\documentclass[
]{ceurart}

\usepackage{listings}
\begin{document}

\copyrightyear{2024}
\copyrightclause{Copyright for this paper by its authors.
  Use permitted under Creative Commons License Attribution 4.0
  International (CC BY 4.0).}

\conference{PhysioCHI: Towards Best Practices for Integrating Physiological Signals in HCI,   May 11, 2024, Honolulu, HI, USA}

\title{Wearable Biosensors for Early Detection and Continuous Monitoring of Postpartum Depression: \\A Human-Centered Approach}


\author[1]{Emma G. {Ricci-De Lucca}}[%
orcid=0000-0003-4150-5294,
email=emmariccid.th@dartmouth.edu,
]
\address[1]{Thayer School of Engineering, Dartmouth College,
  15 Thayer Drive, Hanover, NH, 03755, United States}

\author[1]{Elizabeth L. Murnane}[%
orcid=0000-0003-2089-207X,
email=emurnane@dartmouth.edu,
]

\cortext[1]{Corresponding author.}
\fntext[1]{E.R.D. conceived of the presented idea and wrote the paper. E.M. supervised and advised in the process.}

\begin{abstract}
   Postpartum depression (PPD) is the leading cause of maternal mortality in the United States. However, there are significant gaps in PPD screening, detection, and treatment. A main challenge stems from limitations in existing methods for early PPD diagnosis and subsequent monitoring and intervention. Mobile technology that incorporates interactive interfaces and continuous wearable biosensors holds major potential to augment clinical decision-making and enhance PPD care. A number of compelling research directions exist in effectively developing such systems, including user experience design, sensing accuracy during perinatal stages, clinical integration and interoperability, acceptance and adoption, and privacy and ethics. By focusing on wearable devices that center women and their experiences, this paper discusses such challenges and opportunities at the intersections of HCI, inclusive physiological computing, and women's health.
\end{abstract}


\maketitle

\section{Introduction}

According to the World Health Organization and the Centers for Disease Control and Prevention, achieving global health objectives requires prioritizing the health of women and children --- and yet “maternal mortality is unacceptably high” \cite{WorldHealthOrganization2014DepartmentMortality}. Notwithstanding immense medical and technological progress, more than 700 women die every day globally from preventable causes related to pregnancy and childbirth, with 90\% of those occurring in low- and middle-income countries \cite{WorldHealthOrganization2019TrendsMortality}. Encouragingly, 80\% of the causes of perinatal morbidity and maternal deaths are preventable when recognized early \cite{Berg2005PreventabilityIII, Zuckerwise2017MaternalResponsiveness, trost2024pregnancy}. However, failure to quickly detect and respond to worsening clinical signs is commonly linked to maternal deaths \cite{Bowyer2008TheUK, Knight2015Saving2009-13}. 

\begin{figure}[h]
  \centering
  \includegraphics[width=1.0\linewidth]{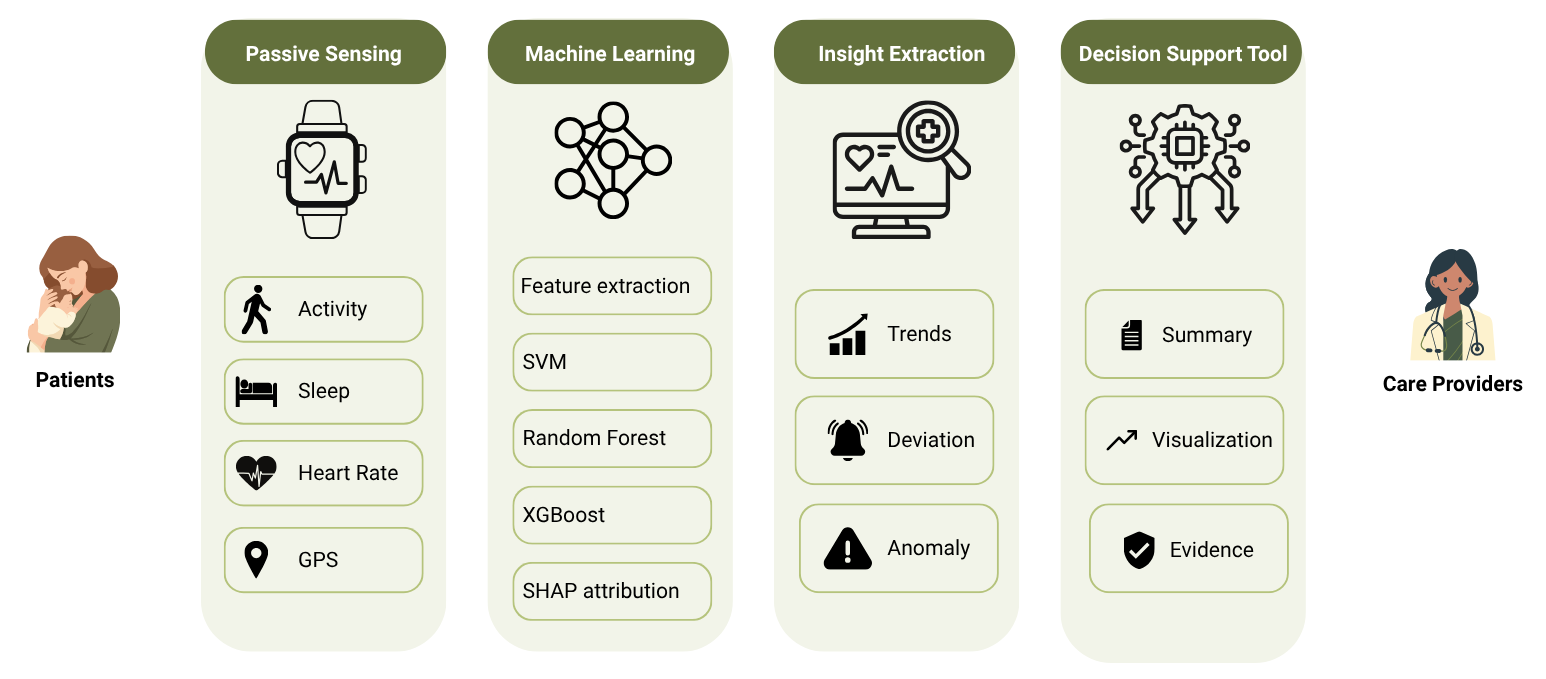}
  \caption{Pipeline for wearable-based postpartum depression detection and care, from passive sensing through machine learning and insight extraction to clinical decision support.}
  \label{fig:pipeline}
\end{figure}

Given wearable monitoring technology supports early detection and intervention, we highlight its promise in addressing maternal health challenges \cite{Lanssens2020DesignDisorders}. Figure~\ref{fig:pipeline} illustrates this vision: passive sensing feeds into machine learning models, which extract clinically relevant insights and surface them through a decision support tool for care providers. In this paper, we argue for a human-centered approach in designing a wearable device that can sense maternal physiological data in order to rapidly identify and intervene in case of deteriorating clinical indicators. 

\section{Maternal mental health crisis}

In the United States, slightly more than half (52\%) of all deaths occur after giving birth and within 42 days of delivery \cite{Declercq2020MaternalPrimer}. The leading underlying cause of pregnancy-related deaths is mental health conditions (23\%) and most notably postpartum depression (PPD) \cite{Petersen2019Vital20132017}. PPD is characterized by intense feelings of sadness, anxiety, and tiredness that can start within 1 to 3 weeks of having a baby \cite{2019PostpartumDepression}. This medical condition can be mistaken for the "baby blues" \cite{mckelvey2018postpartum}, which many new moms experience and is also denoted by mood swings, anxiety, and crying spells; but PPD is a more severe and long-lasting form of depression \cite{2022PostpartumDepression}.

Although postpartum depression is a commonly diagnosed issue following childbirth, there are significant gaps in the screening, detection, and treatment of women suffering from this condition \cite{Patel2012PostpartumReview, Thornicroft2017UndertreatmentCountries}. The repercussions of PPD extend far and wide, with potentially devastating impacts. PPD is viewed as the leading risk factor for maternal suicide and infanticide, making it a serious concern \cite{Fuhr2014ContributionMeta-analysis, Mangla2019MaternalOutcome}. Deaths among mothers that are linked to underlying perinatal mood disorders are seen as preventable with proper awareness and intervention \cite{Trost2021Preventing200817}. Unfortunately, the diagnosis of PPD is currently constrained by the methods employed to identify women who are experiencing this condition. Typically, self-report checklists like the Edinburgh Postnatal Depression Scale (EPDS) \cite{Cox1987DetectionScale} and the Patient Health Questionnaire (PHQ-9) \cite{KroenkeTheMeasure} are utilized to determine which individuals should be included in an intervention program \cite{Singla2017PsychologicalCountries}. However, these approaches have their limitations and there is a need for more comprehensive and accurate methods for both initial detection and ongoing monitoring in order to effectively address and support women suffering from PPD.

Ensuring sufficient care for women after childbirth, which includes postpartum visits, is crucial for the well-being of mothers in the long run and for effectively reducing the rates of maternal deaths \cite{Attanasio2022PostpartumReview}. Research shows that a significant number of women fail to attend the comprehensive postpartum visit, despite its importance for obtaining care for acute issues. According to a systematic review of 88 studies, the average rate of postpartum visit attendance is 72.1\% in the US, yet rates vary substantially, ranging from 24.9\% to 96.5\%, with women from marginalized social groups showing markedly lower rates of postpartum visit attendance \cite{Attanasio2022PostpartumReview}.

\begin{figure}[h]
  \centering
  \includegraphics[width=1.0\linewidth]{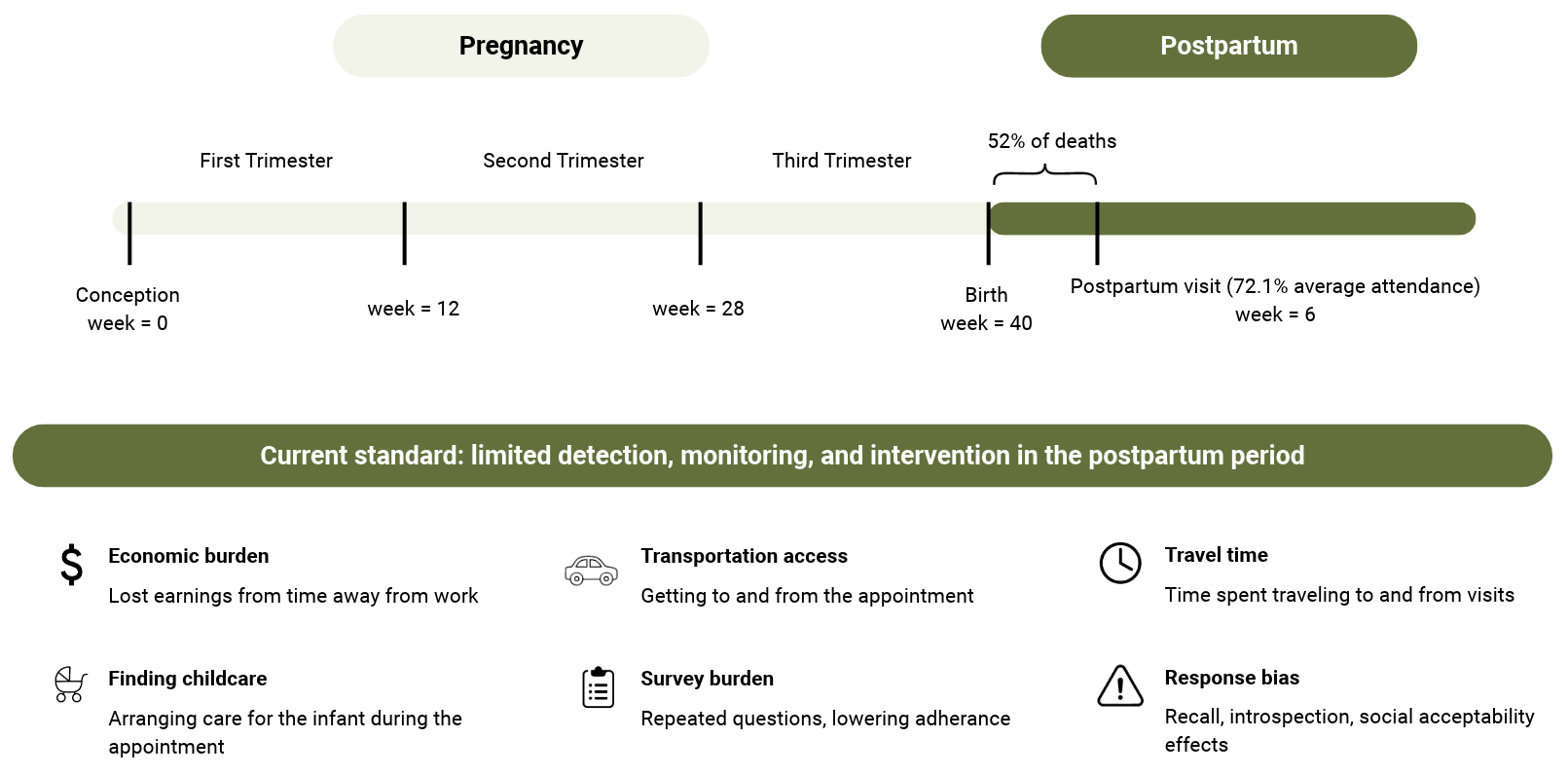}
  \caption{Timing of postpartum depression symptom onset relative to standard postpartum care, illustrating the surveillance gap between birth and the six-week postpartum visit.}
  \label{fig:ppd-timing}
\end{figure}

When considering attendance rates, it is important to recognize that postpartum visits can pose significant economic and time burdens on patients such as time spent traveling to and from appointments and earnings lost from time away from work. While a self-report survey could act as a relatively inexpensive and accessible substitute for in-person visits to examine a patient’s status and PPD symptoms, answering the same set of questions repeatedly can be tiring, time-consuming, and generally burdensome, leading to decreased adherence. Further, self-reporting faces various well-established response biases (e.g., retrospective recall, introspective ability, social acceptability, etc). Figure~\ref{fig:ppd-timing} illustrates this surveillance gap between birth and the standard six-week visit. In light of such challenges, leveraging mobile sensors for the passive collection of PPD indicators, including biosignals, is a promising method to gain understanding and draw inferences on not only psychological but also behavioral and physiological states \cite{hart2022using}. Further, sensor-enabled interactive technology offers the ability to deliver personalized, just-in-time interventions in response to these data and a user's evolving PPD-care needs. 

Current commercialized solutions that address the maternal mental health crisis (such as Mammha, Seven Starling, Babyscripts, Pomelo Care, and Ouma) consist of treatment platforms that offer telehealth and screening services. These services, however, do not leverage clinical data gathered by physiological sensors. While there are a few women’s health solutions that take advantage of wearable devices for data-driven insights (such as Bloomlife, Emagine Solutions Technology, Nuvo, and Oura), none of them integrate the biosignals into a clinical-decision support tool to enable the detection and continuous monitoring of mental health disorders, namely postpartum depression. Thus, there is white space in incorporating physiological sensing with computational power for digital postpartum health applications.

Overall, there is a clear need for a better way to screen for PPD, capture women in the postpartum period, and detect depressive symptoms to prompt timely care and support. Remote monitoring solutions are well-poised to augment traditional postpartum programs, address the shortcomings described above, and reduce the barriers to attending in-person visits and meet women where they are. The current technological landscape presents a unique opportunity to develop digital medicine and integrate it into clinical care. In particular, we focus on wearable devices that can track biosignals pertinent to maternal health and PPD. These devices have the potential to revolutionize healthcare by providing cost-effective and readily accessible remote monitoring, which is especially crucial in regions with limited resources or inadequate healthcare facilities, both within the United States and internationally.

\section{Biosensors for remote monitoring of maternal health}

The American College of Obstetrics and Gynecology (ACOG) has recently published guidelines indicating that mHealth apps have the potential to offer additional clinical support \cite{Auguste2018RecommendationsE141}. In particular, there has been a rise in the availability of commercial apps that focus on pregnancy-related healthcare \cite{Hughson2018TheReview, Tripp2014AnDoctor}. Moreover, the utilization of mobile devices, wearable technologies, and social media provides a vast amount of health data that can be used to accurately evaluate symptoms of mental illnesses like depression \cite{Zarate2022ExploringEvidence}. However, there is sparse academic research on postpartum mental health applications that incorporate biosignals and leverage clinical data, and no such consumer apps are currently available on the market.

\begin{figure}[h]
  \centering
  \includegraphics[width=1.0\linewidth]{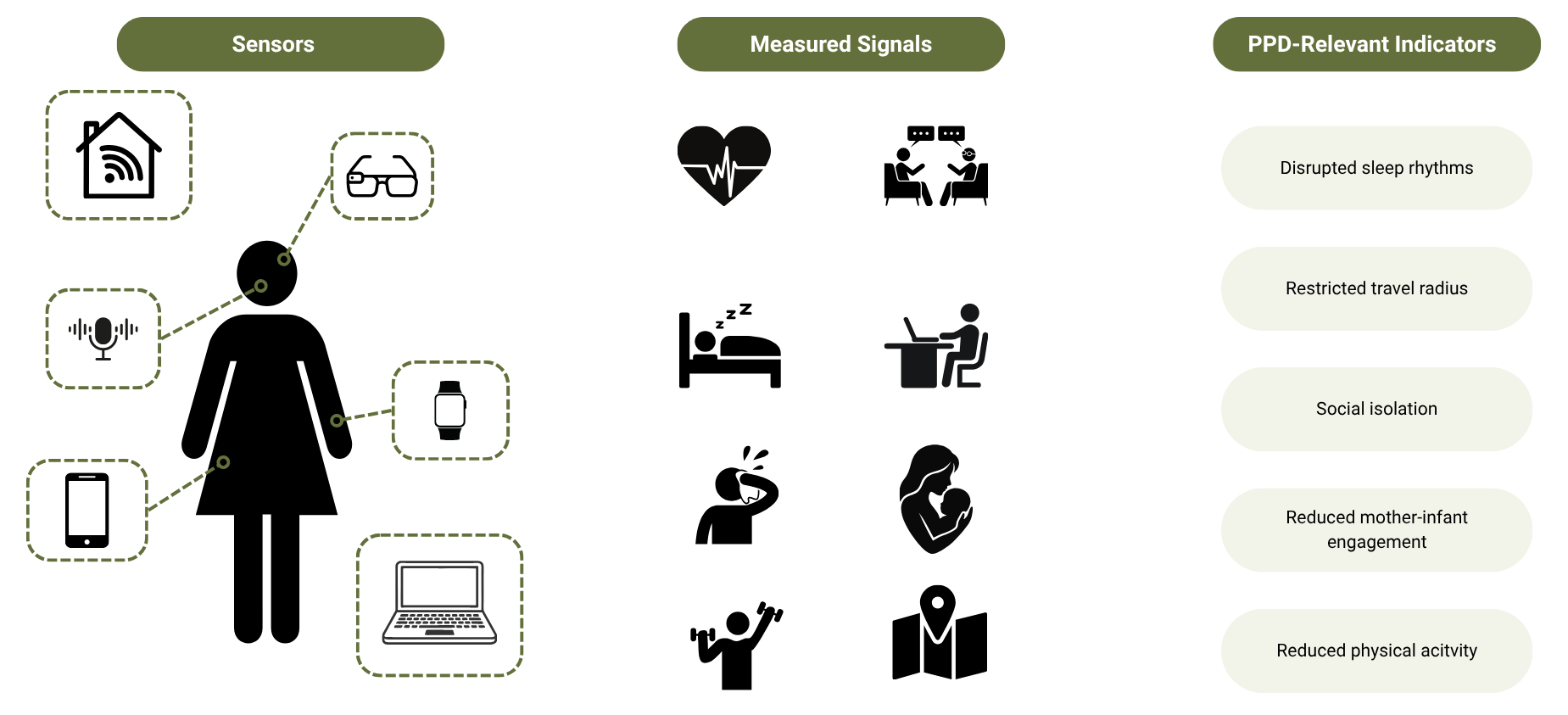}
  \caption{Wearable and ambient sensors capture physiological and behavioral signals linked to known postpartum depression risk indicators.}
  \label{fig:wearables}
\end{figure}

Studies have shown that disrupted sleep and daily rhythms, \cite{Krawczak2016DoPeriod, McBean2015DiurnalWomen} restricted radius of travel \cite{Faherty2017MovementPregnancy}, social isolation \cite{Ganann2016PredictorsChildbirth}, loss of social group membership \cite{Seymour-Smith2017LossMothers}, and limited verbal engagement between mothers and infants are associated with greater depression symptoms \cite{Conners-Burrow2014Low-levelDevelopment, Yan2016SupplementalFunctioning}. Figure~\ref{fig:wearables} summarizes how passive sensors map onto these PPD-relevant indicators. The passive collection of such psychosocial, behavioral, and physiological parameters could therefore offer deep insights into patients’ mental and physical health status, including to identify mothers in need of psychological services and other PPD care. Furthermore, these data can facilitate the personalization of that care and the delivery of in-the-moment support. Consequently, wearable passive sensing technology holds immense potential in capturing struggling mothers early on, delivering treatment, and possibly enhancing outcomes for postpartum depression \cite{Poudyal2019WearableNepal}. Nevertheless, combining these signals can often be ambiguous \cite{Howell2016BiosignalsConductance}, and it is a challenge to integrate them into an embedded, well-designed system that offers both reliable measurement and user-friendly feedback.

A previous study by Poudyal et al. assessed the use of wearable digital sensors to identify young women with depression and to investigate how the data from these passive sensors could inform interventions and treatment \cite{Poudyal2019WearableNepal}. In this study, the authors gathered data from four different inputs to generate the behavioral patterns: physical activity of the mothers from an accelerometer, the daily radius of travel and routine from Global Positioning System (GPS), the time and interactions of the mothers with their infants from Bluetooth beacon proximity data, and the social interactions and auditory ambiance mothers and infants experience with periodic audio recordings. The study found that the incorporation of passive sensing data in the clinical space for adolescent mothers could substantially enhance the outcomes of psychological treatments and facilitate the personalization of care for mothers.
 
A more recent study explored the clinical utility, predictive ability, and adoptability of an app-based screening device for postpartum depression \cite{Krohn2022App-BasedStudy}.  Specifically, Krohn et al. gathered physiological signals such as daily activity (steps taken), heart rate, and sleep by means of an Apple Watch and mobile app, through which women with postpartum depression could log their symptoms. The authors found that using an app-based screening technology paired with a wearable device that passively tracks biosignals is clinically favorable and accepted by both participants and clinicians, thereby having the potential to be a novel solution to increase access to care. Such work demonstrates the benefits of integrating wearable devices and interactive tools into maternal health care, though it has only scratched the surface of important indicators to track, and more work is still necessary to develop mother-facing interfaces to deliver insights and interventions.
 
An additional barrier for women experiencing PPD encompasses the social stigma of seeking proper medical treatment, which engenders reluctance to acknowledge one’s depressive state, as well as a disposition to downplay the severity of one’s symptoms \cite{Krohn2022App-BasedStudy, Dennis2006PostpartumReview}. Hence, passive monitoring of physiological measures can not only avoid stigma-related self-reporting biases, but it may also help women understand that PPD is not "all in their head" but rather a real, treatable medical condition that relates to measurable biological indicators.

Even though there are few innovations within postpartum depression that combine practical mobile and wearable systems for the assessment of mental health status, numerous studies have made use of sensing data to detect non-perinatal depression \cite{chikersal2021detecting, narziev2020stdd, canzian2015trajectories, dagdanpurev2018development, jin2020attention, saeb2016relationship} and other mental illnesses \cite{lanata2014complexity, jiang2018wearable, fletcher2011wearable, tsanas2020objective}. A 2021 comprehensive review found 24 studies constituting wearable-enabled applications, 22 smartphone-enabled applications, and nine ambient sensors in mental health monitoring, none of which addressed postpartum depression or any other female-specific condition \cite{sheikh2021wearable}. While progress is being made in the sphere of general depression, to the best of our knowledge there are limited mobile passive sensing technologies capable of determining occurrence of depression in the context of postpartum women from a combination of smartphone and wearable passive sensing \cite{Krohn2022App-BasedStudy, Poudyal2019WearableNepal}. The literature on smart devices that offer an opportunity to objectively, unobtrusively, and continuously monitor and detect non-perinatal depressive symptoms has laid the groundwork for determining which metrics to measure and how to collect them in order to establish the most comprehensive and ecologically valid data that can provide insights into an individual’s mental state. This rich information concerning the development, testing, and implementation of digital health applications for general depression can begin to elucidate the pertinent physiological signals that warrant measurement and delineate the crucial methodologies for the seamless integration of identified parameters into smart wearable devices, thereby improving efficacy and applicability in the clinical domain.

\section{Challenges and opportunities for physiological computing in the context of perinatal mental health}

\begin{figure}[h]
  \centering
  \includegraphics[width=1.0\linewidth]{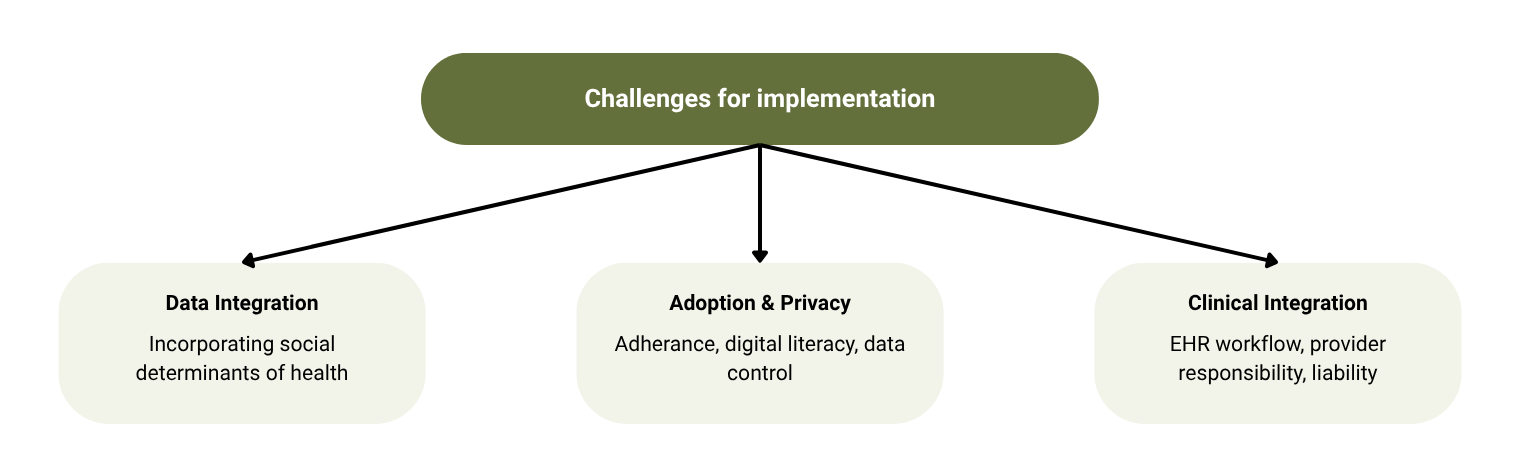}
  \caption{Three interdependent challenge domains in deploying wearable postpartum depression monitoring systems.}
  \label{fig:challenges}
\end{figure}

Biosignals have a substantial (though largely unrealized) potential to equip patients, families, and clinicians with valuable and actionable insights into commonly overlooked biopsychosocial factors in women postpartum. However, the implementation of continuous biosignal monitoring in HCI can present obstacles with regards to interactivity and visual appeal, functionality, clinical integration and interoperability, adoption, and privacy and ethics \cite{Stoyanov2015MobileApps, Stuart2022WearableOpportunities}. The success of such a system requires a harmony in disciplines spanning HCI, psychology, neuroscience, data science, and ethics, which presents a challenge to incorporate into a complex system. Figure~\ref{fig:challenges} outlines these three interdependent domains, each discussed below.

Failure to meet high standards in these domains when creating a wearable device and integrated mobile app can result in low adoption \cite{Inal2020UsabilityReview}, clinical friction, user frustration, or even risk to patients. This danger is even more pronounced among historically marginalized communities, for whom a substandard tool or unreliable assessment functionality could amplify existing health disparities \cite{Miller2017UsabilityPopulations}.

\subsection{Comprehensive data integration}

While existing wearable systems are capable of sophisticated sensor-based monitoring, they often do not account for social determinants of health (SDOH) like race, ethnicity, socioeconomic status, and education, which are crucial predictors of maternal mortality and morbidity. There is room for improvement in better understanding how specific SDOH may impact end-user needs, form factor, and predictive and data analytics capabilities in the context of PPD technology.

Finally, gathering frequent and steady streams of information can provide a more detailed picture of an individual’s behavioral patterns as well as physiological variations over time, which is advantageous in the assessment of symptoms of psychopathologies. Nevertheless, it is important to note that the female body and biometric health parameters, such as heart rate variability, physical activity, and sleep rhythms, undergo enormous changes during pregnancy and the postpartum phase. This necessitates recording and accumulating data from additional weeks during pregnancy, prior to labor, and a need for a more comprehensive model that takes these metrics into account when informing a technological tool. It may also be worthwhile to employ these data to predict patients’ risk for developing depressive symptoms postpartum, such that proactive action can be taken to ensure the patient does not fall through the health care system’s cracks and receives the care that she needs in a timely manner. 

\subsection{Patient adoption, adherence, privacy}

The act of continuously tracking a pregnant woman’s biosignals may worsen existing anxieties, which may lead to a lack of acceptance, adoption, and compliance. Pregnant women may find data collection to be burdensome, especially in the later stages of pregnancy (i.e., the third trimester) and during postpartum, as they are focused and preoccupied with an infant and may struggle to find the time to complete questionnaires or remember to charge and wear a device. Further, swelling and other postpartum changes to the body may impact the comfort of the device. One effective human-centered method for ensuring patients adopt and adhere to a remote patient monitoring device is by incorporating their input and feedback, which involves taking into account their preferences, concerns, and individual needs. By actively involving patients in the initial design processes, healthcare providers can create a more personalized and tailored approach that is more likely to be successful in promoting patient engagement and compliance down the line.

The significant prevalence of mobile technology usage among young individuals from low-income backgrounds in the United States presents a valuable opportunity to utilize it as a crucial resource for providing assistance and support to vulnerable women during their pregnancy and postpartum period \cite{gordon2016participatory}. However, an issue arises when lower levels of digital literacy in some patient populations (e.g., due to limited exposure to technology) causes a lack of understanding and therefore mistrust that can hinder successful adoption and utilization of wearable digital health technologies. To address this, it is necessary to implement education and training programs aimed at improving digital literacy among both patients and healthcare professionals.

Unequal access to digital resources and the financial barriers faced by marginalized communities have a significant impact on their ability to access healthcare services, which risks further widening existing disparities in healthcare. It is therefore important to find ways to reduce the costs or provide financial assistance for wearable digital health technologies. This might be achieved through collaborations between public and private sectors, specifically by forming strategic partnerships between healthcare systems and manufacturers. Such partnerships can help in creating affordable options, offering subsidies, or implementing reimbursement programs that aim to ensure fair and equal access to healthcare resources for all \cite{ginsburg2024key}.

Privacy is another major consideration. Our mental, physical, and emotional physiology is inherently private; and as a result, biodata should also be regarded as highly sensitive and in need of safeguarding. When contemplating the design of a technology in this realm, it is imperative to address aspects such as privacy concerns, individual control over data, fairness and power dynamics, utilization and storage of data. It is also essential to empower individuals with the ability to make an informed choice about sharing their biodata. Furthermore, the complexity of this matter is intensified due to the absence of consistent standards regarding data privacy and security in the healthcare field.

\subsection{Integration into clinical environments}

One of the major hurdles in the clinical implementation of wearable data-driven clinical decision support tools is the large amount of data they generate, which can be overwhelming. As the field advances, efforts are being made to differentiate between situations where raw data should be sent to healthcare systems and cases where summary data would suffice.

A main criticism from healthcare providers about the ever-increasing number of digital solutions is the information overload and complexity of having to keep track of the multitude of complicated, tedious, ancillary dashboards and applications \cite{bersani2020use}. Ideally the continuous and detailed data collected by passive sensing applications would be integrated into Electronic Health Records, which may require transforming the data into standardized formats for seamless exchange between different healthcare systems. The optimal presentation of remote patient monitoring data to healthcare professionals and patients is an ongoing area of research and development \cite{ginsburg2024key}, which can further be informed by ethnographic research and considering users’ constraints, contexts, behaviors, and preferences.

Another integration challenge is determining who should be responsible for monitoring and interpreting this data once it has been communicated. Should it be the obstetrician-gynecologist, primary care physician, doula, or psychiatrist? Taking a human-centered approach involves considering the clinical workflow and interviewing healthcare professionals to determine the most efficient and effective way to utilize these data for both providers and patients. In general, there is currently a lack of clear understanding around the legal responsibilities of physicians utilizing remote patient monitoring with their patients, and the literature is limited in addressing specific medical liability concerns. This lack of clarity can potentially hinder emerging technology’s widespread adoption in clinical practice \cite{patrinos2023medical}. 

\section{Conclusion}

Complex mental health disorders, such as PPD, require meticulous clinical resources and tools to guide decision-making, which is an increasingly crucial aspect in delivering superior care. Unfortunately, while PPD is highly prevalent, most women do not have access to this fine-grained level of personalized treatment. Therefore, it is paramount that more research be carried out on the so-called “fourth trimester” of maternal health, with an emphasis on alleviating growing rates of morbidity and mortality. In particular, interactive technologies that integrate mobile health apps and continuous wearable biosensors that measure physiological parameters can allow for timely identification of the condition so that new mothers can access appropriate treatment and support, thereby helping to improve clinical outcomes and reduce the risk of long-term complications. A user-centered approach is imperative in designing such  tools, prioritizing human factors and clinical implementation as well as interoperability, such that a device can not only provide high quality information but also be broadly inclusive and accessible to populations who will benefit from it the most. 

Relatively speaking, the field of physiological computing is quite young, especially when concerning maternal mental health. Therefore more evidence regarding the effectiveness of approaches and measurement validity is needed, as is ensuring the inclusion of large and diverse samples to avoid further exacerbating health disparities. Currently, the literature lacks research on predicting or detecting perinatal depression on the basis of objective physiological data in the United States while also taking into account social determinants of health. 

As such, the priority should be to investigate factors influencing patient adoption and adherence as well as clinical integration and implementation. The first step to increasing inclusivity and accessibility for these factors would be to take a participatory design approach in conducting thorough user interviews and ethnographic research, while ensuring a diverse and representative pool of participants, to better envisage the end-to-end user journey (including but not limited to needs, pain points, challenges, fears, preferences, etc.) in detail. By focusing on wearable devices that center women and their experiences, our work aims to further global efforts to make medical research more inclusive and accessible.

\section{Acknowledgments}
  This work was supported by the National Science Foundation Research Traineeship, Transformative Research and Graduate Education in Sensor Science, Technology and Innovation (DGE- 2125733).

\bibliography{PPDchi-edited}

\end{document}